\documentclass[12pt,a4paper]{article}
\usepackage{fullpage}
\usepackage{cite}
\usepackage[centertags]{amsmath}
\allowdisplaybreaks[4]
\numberwithin{equation}{section}
\usepackage{amsbsy}
\usepackage{amsfonts}
\usepackage{amssymb}
\usepackage{url}
\usepackage[dvips]{hyperref}
\newcommand{\slashed}[1]{\ensuremath{\rlap{\hskip0.5pt/}{#1}}}
\newcommand{\vet}[1]{\ensuremath{\hskip-1pt\vec{\hskip1pt#1}}}
\newcommand{\newsec}[1]{\section{#1}\label{#1}}

\begin{document}

\begin{flushright}
% \textsf \today}
% \textsf{18 December 2003}
\textsf{25 August 2004}
\\
\textsf{hep-ph/0312256}
\end{flushright}

\vspace{1cm}

\begin{center}
\large
\textbf{Fock States of Flavor Neutrinos are Unphysical}
\normalsize
\\[0.5cm]
\large
Carlo Giunti
\normalsize
\\[0.5cm]
INFN, Sezione di Torino, and Dipartimento di Fisica Teorica,
\\
Universit\`a di Torino,
Via P. Giuria 1, I--10125 Torino, Italy
\\[0.5cm]
\begin{minipage}[t]{0.8\textwidth}
\begin{center}
\textbf{Abstract}
\end{center}
It is shown that
it is possible to construct an infinity of Fock spaces of
flavor neutrinos
depending on arbitrary unphysical mass parameters,
in agreement with the theory of Blasone and Vitiello
in the version proposed by Fujii, Habe and Yabuki.
However,
we show
by \emph{reductio ad absurdum}
that
these flavor neutrino Fock spaces are clever mathematical constructs
without physical relevance,
because
the hypothesis that neutrinos produced or detected in
charged-current weak interaction processes
are described by flavor neutrino Fock states
implies that measurable quantities depend on
the arbitrary unphysical flavor neutrino mass parameters.
\end{minipage}
\end{center}

\begin{flushleft}
\end{flushleft}

%\newpage
%\tableofcontents
%\newpage

\newsec{Introduction}

Neutrino oscillations
\cite{Pontecorvo:1957cp,Pontecorvo:1958qd,Pontecorvo:1968fh}
is one of the main fields of contemporary experimental and theoretical
research in high-energy physics.
The main reason is that neutrino oscillations is a consequence of
neutrino mixing
(see Refs.~\cite{Bilenky:1978nj,Bilenky:1987ty,Valle:1991pk,BGG-review-98,Fisher:1999fb,Bilenkii:2001yh,Gonzalez-Garcia:2002dz,hep-ph/0402025,hep-ph/0310238}
and the the recent review by B. Kayser in Ref.~\cite{Eidelman:2004wy}),
which consists in a mismatch between flavor and mass:
the left-handed flavor neutrino fields $\nu_{\alpha L}$,
with $\alpha=e,\mu,\tau$,
are unitary linear combinations of the massive neutrino fields
$\nu_{kL}$,
\begin{equation}
\nu_{\alpha L}
=
\sum_{k=1}^3 U_{\alpha k} \, \nu_{kL}
\qquad
(\alpha=e,\mu,\tau)
\,,
\label{901}
\end{equation}
where $U$ is the mixing matrix.
Since neutrinos are massless in the Standard Model,
neutrino oscillations represents an open window
on the physics beyond the Standard Model
(see Refs.~\cite{Altarelli:1999gu,Fritzsch:1999ee,Masina:2001pp,hep-ph/0310204})
The theory of neutrino oscillations
has been discussed in many papers
(see Ref.~\cite{Neutrino-Unbound})
and reviewed in
Refs.~\cite{Bilenky:1978nj,Bilenky:1987ty,Boehm:1992nn,CWKim-book,Zralek-oscillations-98,Bilenkii:2001yh,Beuthe:2001rc,Dolgov:2002wy}.

The standard derivation of the neutrino oscillation probability
follows from the description of neutrinos produced or detected
in charged-current weak interaction processes through the flavor neutrino states
\begin{equation}
| \nu_{\alpha} \rangle
=
\sum_{k} U_{\alpha k}^* \, | \nu_k \rangle
\qquad
(\alpha=e,\mu,\tau)
\,,
\label{902}
\end{equation}
where
$| \nu_k \rangle$
is the state of a neutrino with mass $m_k$,
which belongs to the Fock space of the quantized
massive neutrino field $\nu_k$.

It must be noted that the flavor state (\ref{902}) is \emph{not}
a quantum of the flavor field $\nu_{\alpha}$
\cite{Giunti:1992cb}.
Indeed,
one can easily check that
the flavor state (\ref{902}) is not annihilated by the flavor field $\nu_{\alpha}$
if the neutrino masses are taken into account.

In Ref.~\cite{Giunti:1992cb} it was argued that
it is impossible to construct a Fock space of flavor states.
In the proof of this statement it was implicitly assumed that
would-be creation (destruction) operators
of flavor states can be linear combinations of
creation (destruction) operators
of massive states only,
excluding a contribution from destruction (creation) operators
of massive states.
As explained in Section~\ref{Fock space of flavor fields},
this assumption,
although physically reasonable,
is inconsistent with the theory.
It follows that
it is possible to construct a Fock space of flavor states,
as it was first noticed by Blasone and Vitiello (BV)
\cite{Blasone:1995zc} in 1995
and later discussed in several papers by BV with collaborators
\cite{Blasone:1998hf,hep-ph/9907382,Blasone:2002jv},
by Fujii, Habe and Yabuki (FHY)
\cite{Fujii:1998xa,Fujii:2001zv},
by Blasone et al.
\cite{Blasone:2002wp},
by Ji and Mishchenko
\cite{Ji:2002tx},
and other more mathematically oriented authors
\cite{Hannabuss:2000hy}.
Actually,
as shown by FHY \cite{Fujii:1998xa},
there is an infinity of flavor Fock spaces depending
on arbitrary unphysical mass parameters.

It is then necessary to determine if the
flavor Fock states
can describe neutrinos produced or detected in
charged-current weak interaction processes.
As discussed in Section~\ref{Measurable Quantities}
our conclusion is negative,
showing that the flavor Fock spaces
are clever mathematical constructs without physical relevance.
Let us emphasize that
this fact precludes the description of neutrinos in oscillation experiments
through the flavor Fock states,
because
these neutrinos must be produced and detected in
weak interaction processes.

The plane of the paper is as follows.
In Section~\ref{Fock space of flavor fields} we review
the argument presented in Ref.~\cite{Giunti:1992cb}
against a Fock space of flavor states,
we show its inconsistency
and
we explain how an infinity of Fock spaces of flavor states
can be constructed,
obtaining the BV and FHY results through a different way.
In Section~\ref{Measurable Quantities}
we show that the flavor Fock spaces are unphysical
and in Section~\ref{Conclusions}
we summarize our conclusions.

\newsec{Fock space of flavor fields}

There is neutrino mixing if
the mass matrix is not diagonal in the basis
of the flavor neutrino fields $\nu_\alpha(x)$,
where $\alpha=e,\mu,\tau$ is the flavor index.
If we consider, for simplicity,
the mixing of three Dirac neutrinos,
the flavor neutrino fields are related to the
massive neutrino fields
$\nu_{k}(x)$,
where $k=1,2,3$ is the mass index,
by the mixing relation\footnote{
More precisely,
there are two mixing relations for the left and right handed fields
in the basis in which the mass matrix of the
charged lepton fields is diagonal:
$$
\nu_{\alpha L}(x)
=
\sum_{k} U_{\alpha k} \, \nu_{kL}(x)
\,,
\qquad
\nu_{\alpha R}(x)
=
\sum_{k} V_{\alpha k} \, \nu_{kR}(x)
\,,
$$
with the unitary matrices
$U$ and $V$
such that the mass matrix $M$ is diagonalized by the biunitary transformation
$V^{\dagger} M U = M_{\text{diag}}$
(see Ref.~\cite{Bilenky:1987ty}).
However,
since the right-handed fields $\nu_{\alpha R}(x)$
do not participate to weak interactions,
we can define appropriate right-handed flavor fields
$$
\nu'_{\alpha R}(x)
=
\sum_{\beta} \left( U V^{\dagger} \right)_{\alpha \beta} \nu_{\beta R}(x)
=
\sum_{k} U_{\alpha k} \, \nu_{kR}(x)
\,,
$$
such that the flavor fields
$ \nu_\alpha(x) = \nu_{\alpha L}(x) + \nu'_{\alpha R}(x) $
satisfy the mixing relations in Eq.~(\ref{001}).
}
\begin{equation}
\nu_\alpha(x)
=
\sum_{k} U_{\alpha k} \, \nu_{k}(x)
\,,
\label{001}
\end{equation}
where $U$ is the unitary $3\times3$ mixing matrix.

The quantized massive neutrino fields $\nu_{k}(x)$
obey the canonical equal-time anticommutation relations
\begin{equation}
\{
\nu_{k\xi}(t,\vec{x})
\,,\,
\nu_{j\eta}^{\dagger}(t,\vec{y})
\}
=
\delta(\vec{x}-\vec{y})
\,
\delta_{kj}
\,
\delta_{\xi\eta}
\,,
\label{002}
\end{equation}
where $\xi$ and $\eta$ are Dirac indices ($\xi,\eta=1,\ldots,4$),
and
\begin{equation}
\{
\nu_{k\xi}(x)
\,,\,
\nu_{j\eta}(y)
\}
=
\{
\nu_{k\xi}^{\dagger}(x)
\,,\,
\nu_{j\eta}^{\dagger}(y)
\}
=
0
\,.
\label{0021}
\end{equation}
Since the quantized massive neutrino fields
must satisfy the free Dirac equation,
they can be Fourier expanded as
\begin{equation}
\nu_{k}(x)
=
\int
\frac
{ \mathrm{d}\vet{p} }
{ (2\pi)^{3/2} }
\sum_{h=\pm1}
\left[
a_{\nu_{k}}(\vet{p},h)
\,
u_{\nu_{k}}(\vet{p},h)
\,
e^{ - i E_{\nu_{k}} t + i \vet{p} \vet{x} }
+
b_{\nu_{k}}^{\dagger}(\vet{p},h)
\,
v_{\nu_{k}}(\vet{p},h)
\,
e^{ i E_{\nu_{k}} t - i \vet{p} \vet{x} }
\right]
\,,
\label{003}
\end{equation}
where
$E_{\nu_{k}}=\sqrt{\vet{p}^2+ m_{\nu_{k}}^2}$,
$h$ is the helicity,
$u_{\nu_{k}}(\vet{p},h)$
and
$v_{\nu_{k}}(\vet{p},h)$
are the usual four-component spinors in momentum space
such that
\begin{equation}
\left( \slashed{p} - m_{\nu_{k}} \right) u_{\nu_{k}}(\vet{p},h) = 0
\,,
\qquad
\left( \slashed{p} + m_{\nu_{k}} \right) v_{\nu_{k}}(\vet{p},h) = 0
\,,
\label{100}
\end{equation}
for which
we use the BV
normalization \cite{Blasone:1995zc}
\begin{equation}
u_{\nu_{k}}^{\dagger}(\vet{p},h)
\,
u_{\nu_{k}}(\vet{p},h')
=
v_{\nu_{k}}^{\dagger}(\vet{p},h)
\,
v_{\nu_{k}}(\vet{p},h')
=
\delta_{hh'}
\,.
\label{004}
\end{equation}
The following orthogonality and completeness relations are useful:
\begin{equation}
u_{\nu_{k}}^{\dagger}(\vet{p},h)
\,
v_{\nu_{k}}(-\vet{p},h')
=
0
\,,
\label{005}
\end{equation}
\begin{equation}
\sum_h
\left(
u_{\nu_{k}}(\vet{p},h) \, u_{\nu_{k}}^{\dagger}(\vet{p},h)
+
v_{\nu_{k}}(-\vet{p},h) \, v_{\nu_{k}}^{\dagger}(-\vet{p},h)
\right)
=
1
\,.
\label{006}
\end{equation}
Using the orthonormality relations
(\ref{004}) and (\ref{005}),
one can find that
\begin{eqnarray}
&&
a_{\nu_{k}}(\vet{p},h)
=
\int
\frac
{ \mathrm{d}\vet{x} }
{ (2\pi)^{3/2} }
\,
e^{ i E_{\nu_{k}} t - i \vet{p} \vet{x} }
\,
u_{\nu_{k}}^{\dagger}(\vet{p},h)
\,
\nu_{k}(x)
\,,
\label{0071}
\\
&&
b_{\nu_{k}}(\vet{p},h)
=
\int
\frac
{ \mathrm{d}\vet{x} }
{ (2\pi)^{3/2} }
\,
\nu_{k}^{\dagger}(x)
\,
v_{\nu_{k}}(\vet{p},h)
\,
e^{ i E_{\nu_{k}} t - i \vet{p} \vet{x}}
\,.
\label{0072}
\end{eqnarray}
The canonical anticommutation relations
(\ref{002}) and (\ref{0021})
for the massive neutrino fields
imply that
\begin{equation}
\{
a_{\nu_{k}}(\vet{p},h)
\,,\,
a_{\nu_{j}}^{\dagger}(\vet{p}',h')
\}
=
\{
b_{\nu_{k}}(\vet{p},h)
\,,\,
b_{\nu_{j}}^{\dagger}(\vet{p}',h')
\}
=
\delta(\vet{p}-\vet{p}') \, \delta_{hh'} \, \delta_{kj}
\,,
\label{009}
\end{equation}
and
all the other anticommutation relations vanish.
Since these are the canonical anticommutation relations
for fermionic ladder operators,
the operators
$a_{\nu_{k}}^{\dagger}(\vet{p},h)$ and $b_{\nu_{k}}^{\dagger}(\vet{p},h)$
can be interpreted, respectively, as
the one-particle and one-antiparticle creation operators
which allow to construct
the Fock space of massive neutrino states
starting from the vacuum ground state $|0\rangle$.

Let us now consider the flavor fields $\nu_\alpha(x)$.
In order to generate a Fock space of flavor states,
the Fourier expansion of the flavor fields must be written
as
\begin{equation}
\nu_{\alpha}(x)
=
\int
\frac
{ \mathrm{d}\vet{p} }
{ (2\pi)^{3/2} }
\sum_{h=\pm1}
\left[
a_{\nu_{\alpha}}(\vet{p},h)
\,
u_{\nu_{\alpha}}(\vet{p},h)
\,
e^{ - i E_{\nu_{\alpha}} t + i \vet{p} \vet{x} }
+
b_{\nu_{\alpha}}^{\dagger}(\vet{p},h)
\,
v_{\nu_{\alpha}}(\vet{p},h)
\,
e^{ i E_{\nu_{\alpha}} t - i \vet{p} \vet{x} }
\right]
\,,
\label{010}
\end{equation}
where $E_{\nu_{\alpha}}=\sqrt{\vet{p}^2+\widetilde{m}_{\nu_{\alpha}}^2}$
with arbitrary mass parameters $\widetilde{m}_{\nu_{\alpha}}$,
and
the spinors
$u_{\nu_{\alpha}}(\vet{p},h)$ and $v_{\nu_{\alpha}}(\vet{p},h)$
are assumed to satisfy equations analogous to the ones in Eq.~(\ref{100})
\cite{Fujii:1998xa}:
\begin{equation}
\left( \slashed{p} - \widetilde{m}_{\nu_{\alpha}} \right) u_{\nu_{\alpha}}(\vet{p},h) = 0
\,,
\qquad
\left( \slashed{p} + \widetilde{m}_{\nu_{\alpha}} \right) v_{\nu_{\alpha}}(\vet{p},h) = 0
\,.
\label{101}
\end{equation}
Hence, the spinors $u_{\nu_{\alpha}}(\vet{p},h)$ and $v_{\nu_{\alpha}}(\vet{p},h)$
satisfy orthonormality and completeness relations analogous to those in
Eqs.~(\ref{004})--(\ref{006}):
\begin{equation}
u_{\nu_{\alpha}}^{\dagger}(\vet{p},h)
\,
u_{\nu_{\alpha}}(\vet{p},h')
=
v_{\nu_{\alpha}}^{\dagger}(\vet{p},h)
\,
v_{\nu_{\alpha}}(\vet{p},h')
=
\delta_{hh'}
\,,
\label{0041}
\end{equation}
\begin{equation}
u_{\nu_{\alpha}}^{\dagger}(\vet{p},h)
\,
v_{\nu_{\alpha}}(-\vet{p},h')
=
0
\,,
\label{0051}
\end{equation}
\begin{equation}
\sum_h
\left(
u_{\nu_{\alpha}}(\vet{p},h) \, u_{\nu_{\alpha}}^{\dagger}(\vet{p},h)
+
v_{\nu_{\alpha}}(-\vet{p},h) \, v_{\nu_{\alpha}}^{\dagger}(-\vet{p},h)
\right)
=
1
\,.
\label{0061}
\end{equation}

Using Eq.~(\ref{003}),
the mixing relation 
(\ref{001}) allows to write the flavor fields as
\begin{equation}
\nu_\alpha(x)
=
\int
\frac
{ \mathrm{d}\vet{p} }
{ (2\pi)^{3/2} }
\sum_{h=\pm1}
\sum_{k}
U_{\alpha k}
\left[
a_{\nu_{k}}(\vet{p},h)
\,
u_{\nu_{k}}(\vet{p},h)
\,
e^{ - i E_{\nu_{k}} t + i \vet{p} \vet{x} }
+
b_{\nu_{k}}^{\dagger}(\vet{p},h)
\,
v_{\nu_{k}}(\vet{p},h)
\,
e^{ i E_{\nu_{k}} t - i \vet{p} \vet{x} }
\right]
\,.
\label{011}
\end{equation}
Confronting with Eq.~(\ref{010})
and assuming that the would-be destruction (creation) operators of flavor states
are linear combinations of destruction (creation) operators of massive states only,
for the would-be destruction operators of flavor neutrino states
$a_{\nu_{\alpha}}(\vet{p},h)$
we have
\begin{equation}
a_{\nu_{\alpha}}(\vet{p},h)
\,
u_{\nu_{\alpha}}(\vet{p},h)
\,
e^{ - i E_{\nu_{\alpha}} t }
=
\sum_{k}
U_{\alpha k}
\,
a_{\nu_{k}}(\vet{p},h)
\,
u_{\nu_{k}}(\vet{p},h)
\,
e^{ - i E_{\nu_{k}} t }
\,.
\label{012}
\end{equation}
Using the orthonormality relation (\ref{0041})
we obtain
\begin{equation}
a_{\nu_{\alpha}}(\vet{p},h)
=
\sum_{k}
U_{\alpha k}
\,
a_{\nu_{k}}(\vet{p},h)
\Big(
u_{\nu_{\alpha}}^{\dagger}(\vet{p},h)
\,
u_{\nu_{k}}(\vet{p},h)
\Big)
e^{ i \left( E_{\nu_{\alpha}} - E_{\nu_{k}} \right) t }
\,,
\label{013}
\end{equation}
With the help of Eq.~(\ref{009})
one can calculate
the anticommutation relation
\begin{align}
\{
a_{\nu_{\alpha}}(\vet{p},h)
\,,\,
a_{\nu_{\beta}}^{\dagger}(\vet{p}',h')
\}
\null & \null
=
\delta(\vet{p}-\vet{p}') \, \delta_{hh'}
\,
e^{ i \left( E_{\nu_{\alpha}} - E_{\nu_{\beta}} \right) t }
\nonumber
\\
\null & \null
\times
u_{\nu_{\alpha}}^{\dagger}(\vet{p},h)
\left(
\sum_{k}
U_{\alpha k}
\,
U_{\beta k}^*
\,
u_{\nu_{k}}(\vet{p},h)
\,
u_{\nu_{k}}^{\dagger}(\vet{p},h)
\right)
u_{\nu_{\beta}}(\vet{p},h)
\,,
\label{014}
\end{align}
which is not proportional to $\delta_{\alpha\beta}$
because of the $4\times4$ matrix coefficients
$
u_{\nu_{k}}(\vet{p},h)
\,
u_{\nu_{k}}^{\dagger}(\vet{p},h)
$
that prevent the operativeness of the
unitarity relation
$ \displaystyle
\sum_{k}
U_{\alpha k}
\,
U_{\beta k}^*
=
\delta_{\alpha\beta}
$.
A similar derivation applies to the
operators
$b_{\nu_{\alpha}}(\vet{p},h)$.

From these considerations
one can see that the operators
$a_{\nu_{\alpha}}(\vet{p},h)$
and
$b_{\nu_{\alpha}}(\vet{p},h)$
calculated in this way
do not have the properties of fermionic ladder operators.
From similar considerations,
in Ref.~\cite{Giunti:1992cb}
it was concluded that a Fock space of flavor states do not exist.
Let us emphasize again that
this conclusion follows from the assumption that
the would-be destruction (creation) operators of flavor states
are linear combinations of destruction (creation) operators of massive states only.
This is equivalent to assume that the
vacuum of the Fock space of flavor states
is the same as the vacuum of the Fock space of massive states,
because the vacuum of the Fock space of massive states
is obviously annihilated by the operators
$a_{\nu_{\alpha}}(\vet{p},h)$
in Eq.~(\ref{013})
(and by the $b_{\nu_{\alpha}}(\vet{p},h)$ defined in an analogous way).
We think that this is a necessary requirement
for a physical interpretation of Fock space of flavor states,
because there is only one vacuum in the real world.

However,
the condition that
the would-be destruction (creation) operators of flavor states
are linear combinations of destruction (creation) operators of massive states only
is in contradiction with
the Fourier expansion (\ref{010}) of $\nu_{\alpha}(x)$ and
the orthonormality relations of
the spinors
$u_{\nu_{\alpha}}(\vet{p},h)$
and
$v_{\nu_{\alpha}}(\vet{p},h)$
in Eqs.~(\ref{0041})--(\ref{0061}),
which
imply that the operators
$a_{\nu_{\alpha}}(\vet{p},h)$
and
$b_{\nu_{\alpha}}(\vet{p},h)$
are given by relations analogous to those in Eqs.~(\ref{0071}) and (\ref{0072}):
\begin{eqnarray}
&&
a_{\nu_{\alpha}}(\vet{p},h)
=
\int
\frac
{ \mathrm{d}\vet{x} }
{ (2\pi)^{3/2} }
\,
e^{ i E_{\nu_{\alpha}} t - i \vet{p} \vet{x} }
\,
u_{\nu_{\alpha}}^{\dagger}(\vet{p},h)
\,
\nu_{\alpha}(x)
\,,
\label{0151}
\\
&&
b_{\nu_{\alpha}}(\vet{p},h)
=
\int
\frac
{ \mathrm{d}\vet{x} }
{ (2\pi)^{3/2} }
\,
\nu_{\alpha}^{\dagger}(x)
\,
v_{\nu_{\alpha}}(\vet{p},h)
\,
e^{ i E_{\nu_{\alpha}} t - i \vet{p} \vet{x}}
\,.
\label{0152}
\end{eqnarray}
Using the mixing relation (\ref{001})
and the Fourier expansion (\ref{003})
of the massive neutrino fields,
we obtain
\begin{align}
a_{\nu_{\alpha}}(\vet{p},h)
=
e^{ i E_{\nu_{\alpha}} t }
\sum_{k}
U_{\alpha k}
\null & \null
\left[
a_{\nu_{k}}(\vet{p},h)
\Big(
u_{\nu_{\alpha}}^{\dagger}(\vet{p},h)
\,
u_{\nu_{k}}(\vet{p},h)
\Big)
e^{ - i E_{\nu_{k}} t }
\right.
\nonumber
\\
\null & \null
\left. \null
+
b_{\nu_{k}}^{\dagger}(-\vet{p},h)
\Big(
u_{\nu_{\alpha}}^{\dagger}(\vet{p},h)
\,
v_{\nu_{k}}(-\vet{p},h)
\Big)
e^{ i E_{\nu_{k}} t }
\right]
\,,
\label{016}
\\
b_{\nu_{\alpha}}(\vet{p},h)
=
e^{ i E_{\nu_{\alpha}} t }
\sum_{k}
U_{\alpha k}^*
\null & \null
\left[
a_{\nu_{k}}^{\dagger}(-\vet{p},h)
\Big(
u_{\nu_{k}}^{\dagger}(-\vet{p},h)
\,
v_{\nu_{\alpha}}(\vet{p},h)
\Big)
e^{ i E_{\nu_{k}} t }
\right.
\nonumber
\\
\null & \null
\left. \null
+
b_{\nu_{k}}(\vet{p},h)
\Big(
v_{\nu_{k}}^{\dagger}(\vet{p},h)
\,
v_{\nu_{\alpha}}(\vet{p},h)
\Big)
e^{ - i E_{\nu_{k}} t }
\right]
\,.
\label{017}
\end{align}
These relations are identical to those obtained by
FHY \cite{Fujii:1998xa,Fujii:2001zv}
(see also Ref.~\cite{hep-ph/9907382})
through a
generalization of the BV formalism\footnote{
BV
assumed that
$\widetilde{m}_{\nu_{e}}=m_{\nu_{1}}$,
$\widetilde{m}_{\nu_{\mu}}=m_{\nu_{2}}$,
$\widetilde{m}_{\nu_{\tau}}=m_{\nu_{3}}$.
}
\cite{Blasone:1995zc}.
The operators
$a_{\nu_{\alpha}}(\vet{p},h)$
and
$b_{\nu_{\alpha}}(\vet{p},h)$
satisfy the canonical anticommutation relations
\begin{equation}
\{
a_{\nu_{\alpha}}(\vet{p},h)
\,,\,
a_{\nu_{\beta}}^{\dagger}(\vet{p}',h')
\}
=
\{
b_{\nu_{\alpha}}(\vet{p},h)
\,,\,
b_{\nu_{\beta}}^{\dagger}(\vet{p}',h')
\}
=
\delta(\vet{p}-\vet{p}') \, \delta_{hh'} \, \delta_{\alpha\beta}
\,,
\label{019}
\end{equation}
and
all the other anticommutation relations vanish.
Therefore,
the argument presented in Ref.~\cite{Giunti:1992cb}
against a Fock space of flavor states
is inconsistent
and,
as pointed out by BV \cite{Blasone:1995zc},
the operators
$a_{\nu_{\alpha}}^{\dagger}(\vet{p},h)$
and
$b_{\nu_{\alpha}}^{\dagger}(\vet{p},h)$
can be interpreted, respectively, as
the one-particle and one-antiparticle creation operators
which allow to construct
a Fock space of flavor neutrino states
starting from a vacuum ground state.
However,
such vacuum ground state is different from the vacuum ground state
of massive neutrinos,
that we have denoted by $|0\rangle$,
as one can immediately see from the fact that
the operators
$a_{\nu_{\alpha}}(\vet{p},h)$
and
$b_{\nu_{\alpha}}(\vet{p},h)$
in Eqs.~(\ref{016}) and (\ref{017})
do not annihilate $|0\rangle$:
\begin{align}
\null & \null
a_{\nu_{\alpha}}(\vet{p},h)
\,
|0\rangle
=
e^{ i E_{\nu_{\alpha}} t }
\sum_{k}
U_{\alpha k}
\Big(
u_{\nu_{\alpha}}^{\dagger}(\vet{p},h)
\,
v_{\nu_{k}}(-\vet{p},h)
\Big)
e^{ i E_{\nu_{k}} t }
\,
|\bar\nu_{k}(-\vet{p},h)\rangle
\,,
\label{021}
\\
\null & \null
b_{\nu_{\alpha}}(\vet{p},h)
\,
|0\rangle
=
e^{ i E_{\nu_{\alpha}} t }
\sum_{k}
U_{\alpha k}^*
\Big(
u_{\nu_{k}}^{\dagger}(-\vet{p},h)
\,
v_{\nu_{\alpha}}(\vet{p},h)
\Big)
e^{ i E_{\nu_{k}} t }
\,
|\nu_{k}(-\vet{p},h)\rangle
\,.
\label{022}
\end{align}
Therefore,
the vacuum ground state of the flavor neutrino Fock space
is different from
the vacuum ground state of the massive neutrino Fock space
\cite{Blasone:1995zc}.
Actually,
there is an infinity of Fock spaces of flavor neutrinos
depending on the values of the arbitrary
parameters $\widetilde{m}_{\nu_{\alpha}}$
\cite{Fujii:1998xa}.

Let us denote with
$|0_{\{\widetilde{m}\}}\rangle$
the vacuum ground state of the flavor neutrino Fock space
corresponding to a set of values 
of the parameters $\widetilde{m}_{\nu_{\alpha}}$.
In principle we should add a suffix ${\{\widetilde{m}\}}$
also to the operators
$a_{\nu_{\alpha}}(\vet{p},h)$
and
$b_{\nu_{\alpha}}(\vet{p},h)$
in Eqs.~(\ref{016}) and (\ref{017}),
but we refrain from complicating the notation in such way,
being understood that the operators are assumed to act on the
corresponding vacuum state with the same
values 
of the parameters $\widetilde{m}_{\nu_{\alpha}}$.

Having proved the mathematical possibility
to construct Fock spaces of flavor neutrinos,
it is necessary to investigate if
these Fock spaces and their associated vacuums
have any physical relevance.
In Section~\ref{Measurable Quantities}
we will see that the hypothesis that
real flavor neutrinos
produced and detected in charged-current weak interaction processes
are described by flavor
Fock states leads to the absurd result
that the arbitrary mass
parameters $\widetilde{m}_{\nu_{\alpha}}$ are measurable.
Hence,
the only Fock space which describes reality
is the
massive neutrino Fock space
and its vacuum ground state
is the physical vacuum.

This fact may be also clear from the above derivation,
in which we started with the quantization of the massive neutrino fields,
which are the fundamental quantities,
and we defined arbitrarily
the Fourier expansion of the flavor fields in Eq.~(\ref{010}),
through the arbitrary mass parameters $\widetilde{m}_{\nu_{\alpha}}$
and the arbitrary relations
(\ref{101}).
Instead the masses of the massive neutrino fields and the relations (\ref{100})
are not arbitrary,
because they are determined by the Dirac Lagrangian,
which implies free Dirac equations
for the massive neutrino fields.
On the other hand,
the flavor neutrino fields do not satisfy any sort of free Dirac equation,
because neutrino mixing implies that the equations
of the flavor neutrino fields are coupled by the off-diagonal mass terms
in the flavor basis.
Indeed,
using Eqs.~(\ref{101}), (\ref{016}) and (\ref{017})
one can directly check that the flavor fields
$\nu_{\alpha}(x)$ in Eq.~(\ref{010})
do not satisfy a free Dirac equation with mass $\widetilde{m}_{\nu_{\alpha}}$,
because the operators
$a_{\nu_{\alpha}}(\vet{p},h)$
and
$b_{\nu_{\alpha}}(\vet{p},h)$
are time-dependent.
Therefore,
the definition of the spinors $u_{\nu_{\alpha}}(\vet{p},h)$ and $v_{\nu_{\alpha}}(\vet{p},h)$
through Eqs.~(\ref{101})
is completely arbitrary and the mass parameters $\widetilde{m}_{\nu_{\alpha}}$
are unphysical.
Indeed, it has been emphasized by FHY \cite{Fujii:1998xa}
that the mass parameters $\widetilde{m}_{\nu_{\alpha}}$
should disappear in all measurable quantities.
Since the Fourier expansion of the flavor fields in Eq.~(\ref{010})
is an arbitrary mathematical construct,
we are not surprised by the fact that the
corresponding Fock space of flavor neutrinos has no physical relevance.

\newsec{Measurable Quantities}

In Ref.~\cite{Blasone:1995zc} BV
define the flavor one-neutrino state as
\begin{equation}
|\nu_{\alpha}(\vet{p},h)\rangle
=
a_{\nu_{\alpha}}^{\dagger}(\vet{p},h) \, |0\rangle
\,,
\label{111}
\end{equation}
whereas in Refs.~\cite{Blasone:1998hf,hep-ph/9907382,Blasone:2002jv}
they adopt the definition
\begin{equation}
|\nu_{\alpha}(\vet{p},h)\rangle
=
a_{\nu_{\alpha}}^{\dagger}(\vet{p},h) \, |0_{\{\widetilde{m}\}}\rangle
\,,
\label{112}
\end{equation}
whose motivations are explained in Ref.~\cite{Blasone:1998hf}.
It seems to us that it is obvious that
the definition (\ref{112}) is the correct one
from the point of view of someone which believes that
the Fock space of flavor states describes reality,
because the states in Eq.~(\ref{112})
belong to such Fock space,
whereas the states in Eq.~(\ref{111})
are time-dependent superpositions of states belonging to the Fock space of massive neutrinos,
if $a_{\nu_{\alpha}}(\vet{p},h)$ is interpreted according to Eq.~(\ref{016}).

In this Section we show that the interpretation of the definition
(\ref{112})
as a physical state
describing a flavor neutrino produced or detected in a
charged-current weak interaction process
leads to the absurd result that the unphysical arbitrary parameters
$\widetilde{m}_{\nu_{\alpha}}$ are measurable.
This does not mean that the definition
(\ref{111})
is any better,
as we will see in the following.

Let us consider the simplest case of the pion decay process
\begin{equation}
\pi^{+} \to \mu^{+} + \nu_{\mu}
\,.
\label{121}
\end{equation}
If the flavor one-neutrino states are real,
the outgoing muon neutrino in Eq.~(\ref{121})
is described by the state
\begin{equation}
|\nu_{\mu}(\vet{p},h)\rangle
=
a_{\mu}^{\dagger}(\vet{p},h) \, |0_{\{\widetilde{m}\}}\rangle
\,.
\label{122}
\end{equation}
The amplitude of the decay is given by
\begin{equation}
\mathcal{A}
=
\langle \mu^{+}(\vet{p}_{\mu},h_{\mu}) , \nu_{\mu}(\vet{p},h) |
- i \int \mathrm{d}^{4}x \, \mathcal{H}_{\text{I}}(x)
| \pi^{+}(\vet{p}_{\pi}) , 0_{\{\widetilde{m}\}} \rangle
\,,
\label{123}
\end{equation}
where we have written explicitly
the vacuum flavor state just to make clear that it is assumed to
correspond to the physical vacuum.
The effective interaction Hamiltonian
$\mathcal{H}_{\text{I}}(x)$
is given by
\begin{equation}
\mathcal{H}_{\text{I}}(x)
=
\frac{G_{\text{F}}}{{\sqrt{2}}}
\,
\overline{\nu_{\mu}}(x)
\,
\gamma^{\rho} \left( 1 - \gamma_5 \right)
\,
\mu(x)
\,
J_{\rho}(x)
\,,
\label{124}
\end{equation}
where
$G_{\text{F}}$
is the Fermi constant
and
$J_{\rho}(x)$
is the hadronic weak current,
whose matrix element is given by
\begin{equation}
\langle 0 | J_{\rho}(x) | \pi^{+}(\vet{p}_{\pi}) \rangle
=
i \, \vet{p}_{\pi\rho} \, f_{\pi} \, \cos\vartheta_{\text{C}} \, e^{-ip_{\pi}x}
\,,
\label{125}
\end{equation}
where
$f_{\pi}$ is the pion decay constant
and
$\vartheta_{\text{C}}$
is the Cabibbo angle
(see, for example, Ref.~\cite{Bilenky:1995zq}).
Using Eqs.~(\ref{010}), (\ref{019}) and (\ref{122}) we obtain
\begin{equation}
\mathcal{A}
=
2 \pi
\,
\frac{G_{\text{F}}}{{\sqrt{2}}}
\,
\vet{p}_{\pi\rho} \, f_{\pi} \, \cos\vartheta_{\text{C}}
\,
\delta^4(p_{\pi}-p_{\mu}-p)
\,
\overline{u_{\nu_{\mu}}}(\vet{p},h)
\,
\gamma^{\rho} \left( 1 - \gamma_5 \right)
\,
v_{\mu}(\vet{p}_{\mu},h_{\mu})
\,.
\label{126}
\end{equation}
It is clear that if this expression were correct
the arbitrary unphysical mass parameter $\widetilde{m}_{\nu_{\mu}}$
would be a measurable quantity,
because the energy of the muon neutrino is
$ E_{\nu_{\mu}} = \sqrt{\vet{p}+\widetilde{m}_{\nu_{\mu}}^2}$.
Since
the unphysical mass parameter $\widetilde{m}_{\nu_{\mu}}$
enters in the energy-conservation delta function and in the spinor
$u_{\nu_{\mu}}(\vet{p},h)$,
it determines
the measurable four-momentum of the muon through energy-momentum conservation
and the measurable decay rate of the pion.
Hence,
we conclude that the state (\ref{122})
is unphysical.

Considering other charged-current weak interaction processes one can
rule out the physical relevance of the states (\ref{112})
for all flavors $\alpha$.

The definition
(\ref{111}),
whatever its meaning,
does not lead to anything better.
In this case
the amplitude of the pion decay (\ref{121})
is given by
\begin{equation}
\mathcal{A}
=
\langle \mu^{+}(\vet{p}_{\mu},h_{\mu}) , \nu_{\mu}(\vet{p},h) |
- i \int \mathrm{d}^{4}x \, \mathcal{H}_{\text{I}}(x)
| \pi^{+}(\vet{p}_{\pi}) , 0 \rangle
\,,
\label{127}
\end{equation}
where we have written explicitly
the vacuum state of the massive neutrino Fock space
in order to make clear that it is assumed to
correspond to the physical vacuum,
and
\begin{equation}
|\nu_{\mu}(\vet{p},h)\rangle
=
a_{\nu_{\mu}}^{\dagger}(\vet{p},h) \, |0\rangle
\,.
\label{128}
\end{equation}
Using Eqs.~(\ref{009}), (\ref{011}) and (\ref{016}),
for the matrix element of the neutrino field
we obtain
\begin{equation}
\langle \nu_{\mu}(\vet{p},h) | \overline{\nu_{\mu}}(x) | 0 \rangle
=
\frac{1}{(2\pi)^{3/2}}
\,
e^{ i E_{\nu_{\mu}} t - i \vet{p} \vet{x} }
\sum_{k}
|U_{\mu k}|^2
\Big(
u_{\nu_{\mu}}^{\dagger}(\vet{p},h)
\,
u_{\nu_{k}}(\vet{p},h)
\Big)
\overline{u_{\nu_{k}}}(\vet{p},h)
\,,
\label{129}
\end{equation}
which implies that again
both energy-momentum conservation and
the pion decay rate depend on the unphysical mass $\widetilde{m}_{\nu_{\mu}}$.

Summarizing,
we have shown that both the definitions
(\ref{111}) and (\ref{112})
adopted by BV, FHY and others
cannot correspond to a physical flavor neutrino state
because they would imply that the arbitrary unphysical parameters $\widetilde{m}_{\nu_{\alpha}}$
are measurable\footnote{
Let us notice that also the arbitrary BV
assumption
$\widetilde{m}_{\nu_{e}}=m_{\nu_{1}}$,
$\widetilde{m}_{\nu_{\mu}}=m_{\nu_{2}}$,
$\widetilde{m}_{\nu_{\tau}}=m_{\nu_{3}}$
does not lead to acceptable results.
For example,
it would imply that the pion decay process (\ref{121})
depends only on the neutrino mass $m_{\nu_{2}}$ if the definition (\ref{112}) is adopted.
On the other hand,
using the definition (\ref{111})
one obtains that energy-momentum conservation depends only on $m_{\nu_{2}}$,
although all the neutrino masses contribute in a complicated way to the decay rate.
}.
Since the introduction of
these unphysical parameters is necessary
for the construction of a Fock space of flavor neutrinos,
we conclude that such Fock spaces are only mathematical
constructs, without physical relevance.

Let us emphasize that
the unacceptable results obtained in this Section
are an unavoidable consequence
of the hypothesis that the flavor Fock space is real,
which means that flavor neutrinos are described
by flavor Fock states.
In this case it is not allowed to use the flavor Fock states
for some calculations
(for example neutrino oscillations)
and the massive Fock states for other calculations
(for example pion decay),
all of
which involve neutrinos created or detected in charged-current
weak interactions\footnote{
Since we have shown that the flavor Fock space
is unphysical,
there is no need to discuss neutral-current weak processes.
}.
The obvious reason is that
the flavor Fock states, if real,
are just the states which describe the neutrinos produced and detected in any
charged-current
weak interaction process,
including those operating in neutrino oscillation experiments.
In the calculation of these processes
the flavor neutrino Fock states would have the same relevance
as the Fock states of all other particles.

The correct way to calculate
decay rates (as well as other processes) taking into account neutrino masses and mixing
has been discussed in Refs.~\cite{Shrock:1980vy,Shrock:1981ct,Shrock:1981wq}.
It is based on the fact that
the massive neutrinos
have definite kinematical properties and constitute the
possible orthogonal asymptotic states of the decay.
In other words,
each decay in a massive neutrino constitutes a possible
decay channel and the total decay probability
is the sum of the decay probabilities in the different massive neutrinos
$\nu_k$
weighted by the squared absolute value
of the element of the mixing matrix that weights the
contribution of $\nu_k$
to the charged-current weak interaction Hamiltonian.
The description of
neutrinos produced or detected in charged-current weak interaction processes
through the standard flavor neutrino states (\ref{902})
leads to the same result \cite{hep-ph/0402217}.
Hence,
the standard flavor neutrino states (\ref{902})
can be used to describe in a consistent framework
neutrino interactions and oscillations
in neutrino oscillation experiments.

\newsec{Conclusions}

We have shown that the argument
presented in Ref.~\cite{Giunti:1992cb}
against the existence of a Fock space of flavor neutrinos
is inconsistent.
Hence,
we agree with BV, FHY and others
\cite{Blasone:1995zc,Blasone:1998hf,hep-ph/9907382,Blasone:2002jv,Fujii:1998xa,Fujii:2001zv,Blasone:2002wp,Ji:2002tx,Hannabuss:2000hy}
that
it is possible to construct a Fock space of flavor neutrinos.
However,
there is an infinity of such
Fock spaces of flavor neutrinos
depending on the values of arbitrary unphysical mass parameters
\cite{Fujii:1998xa}.
We have shown that the hypothesis that the flavor Fock states
describe real flavor neutrinos
produced or detected in weak interaction charged-current
processes
leads to the absurd consequence that
the arbitrary unphysical mass parameters
are measurable quantities.
In particular,
the flavor Fock states
are inadequate for the description of flavor neutrinos
in oscillations experiments,
because these flavor neutrinos
are produced and detected through weak interaction
charged-current processes.
Therefore, we conclude that the
Fock spaces of flavor neutrinos
are ingenious mathematical constructs
without physical relevance.

\end{document}